\documentclass[conference]{IEEEtran}
\IEEEoverridecommandlockouts
\usepackage{algorithm,epsfig,cite,stfloats,bm,bbm,amssymb,amsmath,color, subfigure,graphics,extpfeil,amsthm, caption, microtype}
\usepackage{multirow}

\usepackage[noend]{algorithmic}
\def\BibTeX{{\rm B\kern-.05em{\sc i\kern-.025em b}\kern-.08em
    T\kern-.1667em\lower.7ex\hbox{E}\kern-.125emX}}
\begin{document}

\title{Resource Slicing with Cross-Cell Coordination in Satellite-Terrestrial Integrated Networks}
\author{\IEEEauthorblockN{Mingcheng He$^*$, Huaqing Wu$^\dag$, Conghao Zhou$^*$, and Xuemin (Sherman) Shen$^*$}
\IEEEauthorblockA{$^*$Department of Electrical and Computer Engineering, University of Waterloo, Canada}
\IEEEauthorblockA{$^\dag$Department of Electrical and Software Engineering, University of Calgary, Canada}
\IEEEauthorblockA{E-mails: \{m64he, c89zhou, sshen\}@uwaterloo.ca, huaqing.wu1@ucalgary.com}
}


\maketitle
\begin{abstract}
Satellite-terrestrial integrated networks (STIN) are envisioned as a promising architecture for ubiquitous network connections to support diversified services. 
In this paper, we propose a novel resource slicing scheme with cross-cell coordination in STIN to satisfy distinct service delay requirements and efficient resource usage.
To address the challenges posed by spatiotemporal dynamics in service demands and satellite mobility, we formulate the resource slicing problem into a long-term optimization problem and propose a distributed resource slicing (DRS) scheme for scalable and flexible resource management across different cells.
Specifically, a hybrid data-model co-driven approach is developed, including an asynchronous multi-agent reinforcement learning-based algorithm to determine the optimal satellite set serving each cell and a distributed optimization-based algorithm to make the resource reservation decisions for each slice. 
Simulation results demonstrate that the proposed scheme outperforms benchmark methods in terms of resource usage and delay performance.
\end{abstract}
\setlength{\textfloatsep}{2mm}

\section{Introduction}

The next-generation networks are envisioned to support a wide range of emerging services anytime and anywhere \cite{cheng20216g}.
Low Earth orbit (LEO) satellites, thanks to their large communication coverage and global deployment, show great potential in assisting the existing terrestrial networks in providing ubiquitous connectivity and seamless service provision, especially in remote areas.
Therefore, by cooperatively leveraging the complementary advantages of satellite and terrestrial networks, satellite-terrestrial integrated networks (STIN) emerge as a promising network architecture to facilitate services with various quality of service (QoS) requirements in the 6G era.
This encompasses a wide range of network-reliant services, such as vehicular-to-everything (V2X) applications in vehicular networks from delay-sensitive environment sensing services with stringent delay requirements to delay-tolerant services like high-definition (HD) map distribution and multimedia streaming requiring high throughput \cite{5gaa2020c}.

\begin{figure}
    \centering
    \includegraphics[width=0.36\textwidth]{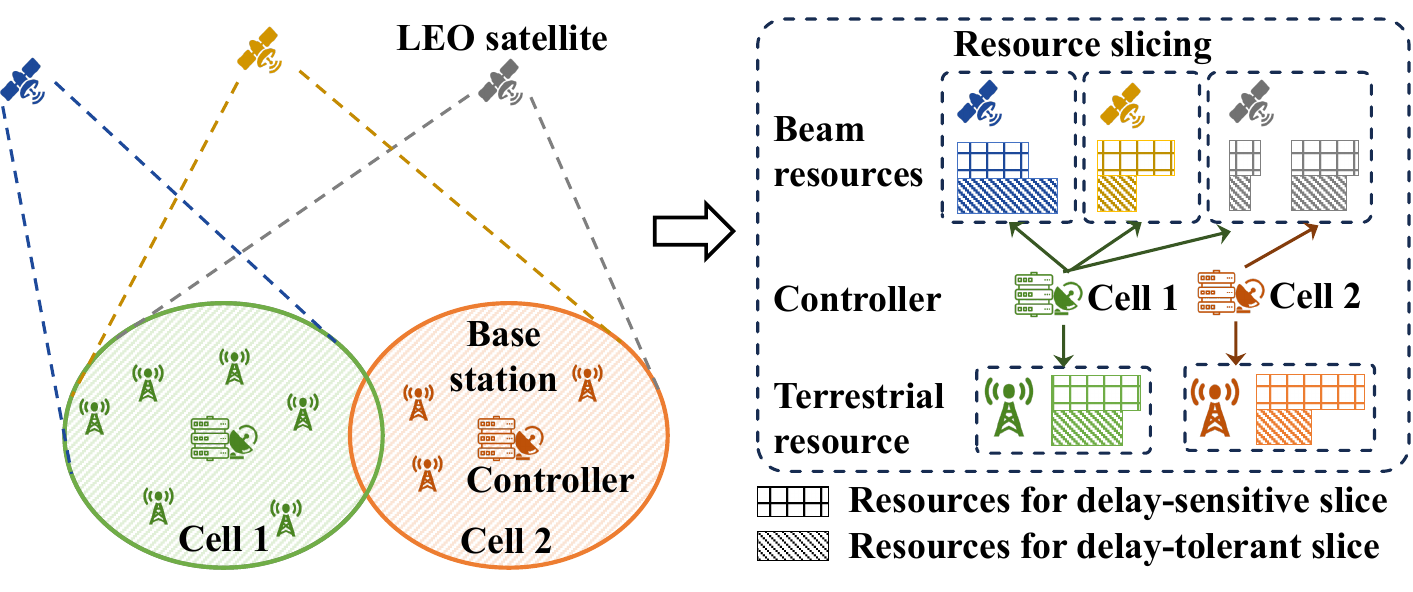}
    \vspace{-1mm}
    \caption{Resource slicing in STIN.}
    \label{fig:architecture}
    \vspace{-2mm}
\end{figure}

To accommodate multifarious services, resource slicing, a significant technique proposed for 5G networks, will continue to play a pivotal role in enabling service differentiation and QoS guarantee in STIN.
Specifically, resource slicing enables the construction of multiple virtual networks, referred to as slices, on top of the shared physical network infrastructure to realize resource isolation while satisfying different service-level agreements.
To maintain the effectiveness of slices, ensuring that they can satisfy QoS requirements in dynamic network conditions with efficient resource utilization, it is crucial to make periodic adjustments to slices at specific intervals referred to as slicing windows. 
In STIN, the terrestrial network consists of multiple cells, with the area of each cell being equal to the coverage area of a satellite beam.  
Within each cell, multiple slices can be created to support different services, and there should be a controller responsible for managing both the terrestrial and satellite resources for different slices, as shown in Fig. \ref{fig:architecture}. 
Moreover, with the large coverage of LEO satellites, resources from each satellite can be shared across different terrestrial cells, introducing the concept of cross-cell coordination of communication resources.

There exist some research efforts on resource slicing in STIN.
An architecture of space-air-ground integrated networks is proposed in \cite{wu2020resource}, where an SDN controller has been designed to help resource slicing decisions. 
In \cite{de2020qos}, a network slicing solution has been presented in STIN to satisfy different grades of QoS in dynamic scenarios for minimizing the latency for resource allocation.
Moreover, a learning-based resource management method has been proposed for STIN in \cite{wu2021learning} combining both resource slicing and scheduling to support two different services to minimize the long-term system cost.
Despite these research endeavors, resource slicing with cross-cell coordination remains an open problem. \looseness=-1

To investigate the resource slicing problem with cross-cell coordination in STIN, three challenges need to be addressed.
First, each satellite has large coverage, which can serve different terrestrial cells using multiple beams by adopting the beam-hopping technique to reuse limited beam resources \cite{lin2022multi}, leading to resource competition among these cells. 
In addition, the dense satellite deployment enables multiple satellites to concurrently serve the same terrestrial cell with frequency reuse or beamforming techniques \cite{peng2022integrating}. 
In this case, implementing a centralized resource slicing approach to derive the optimal decisions will encounter formidable complexity since it requires extensive interactions between different controllers and satellites.
Second, the spatiotemporal dynamics in service demands across different slices and cells result in different lengths of slicing windows as well as asynchronous decision-making in each cell, which further complicates the decision-making process.
Third, due to the mobility of satellites, the amount of resources available for each cell is time-varying and distinct, i.e., for each cell, the number of satellites serving the cell may vary over time and the connection duration to each satellite may be different, posing a challenge to conventional slicing methods designed for cases with fixed resources. \looseness=-1

In this paper, we propose a distributed resource slicing (DRS) scheme with cross-cell coordination in STIN to support scalable and flexible resource management in each cell.
Our objective is to reserve resources from terrestrial and satellite networks for each slice to minimize the overall cost related to resource usage and delay performance.
Specifically, a hybrid data-model co-driven approach is developed to find the optimal solutions in a less time-consuming manner, where an asynchronous multi-agent proximal policy optimization (AMAPPO)-based algorithm is presented to select satellites serving each cell in each slicing window, and a distributed optimization-based algorithm is employed to determine the resource reservation decisions for each slice.
Extensive simulations demonstrate that the proposed scheme can approach the optimal benchmark in terms of resource usage and delay performance with less training time. \looseness=-1

\vspace{-1.5mm}
\section{System Model and Problem Formulation}
\vspace{-1mm}
\subsection{Network Scenario}
\vspace{-1mm}

As shown in Fig. \ref{fig:architecture}, we consider an STIN scenario where the terrestrial network consists of multiple cells, with the area of each cell being equal to the coverage area of a satellite beam. 
The service demand in each terrestrial cell can be served by either terrestrial base stations or LEO satellites.
Let $\mathcal{N} = \{1,2,...,N\}$ denote the set of terrestrial cells, where a controller is deployed in each cell for managing communication resources for different services in the cell.
Denote the set of satellites within a constellation by $\mathcal{S}$. 
Each satellite has $K$ beams ($K < N$), with each beam covering one terrestrial cell.
The available communication bandwidth in each beam is denoted by $B^\mathrm{S}$.
Moreover, there are $A_n$ terrestrial base stations deployed in cell $n \in \mathcal{N}$ where each base station can provide the communication resources with bandwidth $B^\mathrm{T}$.
In this paper, we conceptualize the provisioning of terrestrial communication resources within a cell through a virtual terrestrial access point.
Note that different spectrum frequencies are used for different satellites and terrestrial communications in each cell to avoid co-channel interference. 

Consider a time-slotted system where each time slot $t \in \{1,2,...,T\}$ has the same duration, denoted by $\tau$.
Due to the mobility of satellites, the set of accessible satellites for cell $n$ in time slot $t$ is time-varying, denoted by $\mathcal{S}^{\mathrm{ts}}_{n, t} \subset \mathcal{S}$.
In this case, let $\mathcal{AP}^\mathrm{ts}_{n,t} = \{0\} \cup \mathcal{S}^\mathrm{ts}_{n,t}$ denote the set of all access points being able to serve cell $n$ in time slot $t$, where $ap = 0$ indicates the terrestrial access point.
Let $a_{n, ap, t}$ denote whether access point $ap$ can serve cell $n$ in time slot $t$ or not, where $a_{n, ap, t} = 1$ indicates that access point $ap$ can serve users in cell $n$; otherwise, $a_{n, ap, t} = 0$.

In this paper, the supported services can be classified into two categories: delay-sensitive services with specific delay and reliability requirements and delay-tolerant services without delay requirements.
Two slices, indexed by $l=1$ and $l=2$, are created for the delay-sensitive services and delay-tolerant services, respectively.
Consider the service demand for each service follows a non-homogeneous Poisson Process (NHPP), i.e., the intensity of the Poisson process varies over time slots.
Let $\lambda_{n,l,t}$ denote the service demand intensity of slice $l$ in cell $n$ in time slot $t$.
Specifically, due to the potential intermittent network connection from terrestrial networks in each cell, the service demand can be represented by $\lambda_{n,l,t} = \lambda^\mathrm{C}_{n,l,t} + \lambda^\mathrm{NC}_{n,l,t}$, where $\lambda^\mathrm{C}_{n,l,t}$ is the service demand intensity within the coverage of terrestrial networks, and $\lambda^\mathrm{NC}_{n,l,t}$ is the service demand only being able to be served by satellite networks.
Due to the variation in service demands, the resource slicing decisions should be periodically adjusted to guarantee the effectiveness of resource management, where the interval of the adjustment is referred to as the slicing window.
By analyzing historical data using AI-based prediction methods, in this paper, we assume that the controller can predict the service demand of each slice for the following $W_\mathrm{max}$ time slots.
Therefore, the slicing window length in each terrestrial cell should be determined according to the service demand variation.
Let $\mathcal{W}_{n}^i = \{t_{n, i}, ..., t_{n, i} + w_{n,i} - 1\}$ represent the corresponding time slot set for $i$-th slicing window of cell $n$, where $t_{n, i}$ is the initial time slot of the slicing window, and $w_{n,i} \in \{1, ..., W_\mathrm{max}\}$ is the slicing window length.
Let $\mathcal{AP}_{n,i} \!=\! \bigcup_{t \in \mathcal{W}_{n}^i} \! \mathcal{AP}^\mathrm{ts}_{n,t}$ represents the set of access points that can serve cell $n$ within slicing window $i$.

\vspace{-1.5mm}
\subsection{Communication Model}
\vspace{-0.5mm}

For resource slicing in each slicing window, we aim to reserve communication resources from different access points to each slice to satisfy different service requirements. 
Let $b^{n, ap}_{l, t}$ denote the ratio of reserved resources in access point $ap$ for slice $l$ in cell $n$ at time slot $t$.
Specifically, if $ap=0$, this refers to the ratio of terrestrial resources in cell $n$ reserved for the slice, otherwise, it indicates the ratio of the reserved resources in a beam of $ap$.
Once the resources are reserved, they cannot be adjusted within the slicing window, given by
\vspace{-2mm}
\begin{equation}
    b^{n, ap}_{l, t+1} = b^{n, ap}_{l, t}, \forall t, t+1 \in \mathcal{W}_{n}^i, a_{n, ap, t} = a_{n, ap, t+1} = 1.  \label{op_equal}
    \vspace{-2mm}
\end{equation}

Due to the time-varying nature of stochastic service demands, it is difficult to satisfy the deterministic QoS requirement for every network slice. 
To model the statistical delay performance, effective capacity emerges as a valuable approach \cite{wu2003effective}. 
Specifically, it quantifies the maximum arrival rate that can be accommodated by reserved resources for each slice, facilitating the evaluation of delay-bounded violation probability that the overall delay $D^{n,ap}_{l,t}$ exceeds the delay bound $D_{n, ap, l, t}^{\mathrm{B}}$, presented as \cite{khalek2015delay}
\vspace{-2.5mm}
\begin{equation}
    P(D^{n,ap}_{l,t} \geq D_{n, ap, l, t}^{\mathrm{B}}) \approx e^{-\theta C(R^{n, ap}_{l, t}, \theta) D_{n, ap, l, t}^{\mathrm{B}}},
    \label{eq:delay_violation}
    \vspace{-1.5mm}
\end{equation}
where $C(\cdot, \theta)$ is the effective capacity, and $\theta$ is the QoS exponent representing the exponential decay rate of the QoS violation probability according to the large derivation theory \cite{wu2003effective}, in which the value of $\theta$ is assumed to be the same for queues in each access point.
For a long-term planning problem, we consider the service rate $R^{n, ap}_{l, t}$ to be mainly affected by the average communication distance between users and access points.
In this case, the packet buffer at each access point for each slice can be modeled as an M/D/1 queueing system in each time slot, where the service rate counting in packets in the queue can be given by
\vspace{-1.25mm}
\begin{equation}
    R^{n, ap}_{l, t} = \frac{b^{n, ap}_{l, t} B_{ap}}{L_{l}} \log_2(1 + \frac{P_{ap}|h_{n,ap,t}|^2}{\sigma^2}),
    \vspace{-1.5mm}
\end{equation}
where $L_{l}$ is the packet size for slice $l$, $P_{ap}$ is the transmission power of $ap$, $h_{n,ap,t}$ is the channel gain of the transmission between users in cell $n$ and access point $ap$, which can be presented as $h_{n,ap,t} = (d_{n, ap, t})^{-\delta_{ap}}$, with the pathloss exponent $\delta_{ap}$ of access point $ap$ and the distance $d_{n, ap, t}$ between the central point of cell $n$ and access point $ap$ in time slot $t$.
Denote the entire bandwidth of access point $ap$ by $B_{ap}$. When $ap = 0$, $B_{ap} = B^\mathrm{T}$ for analyzing effective capacity in each base station to determine the delay-bounded violation probability served by terrestrial networks, otherwise, $B_{ap} = B^\mathrm{S}$. 
Moreover, $D_{n, ap, l, t}^{\mathrm{B}} = D_{l}^{\mathrm{B}} - \mathbbm{1}_{ap \in \mathcal{S}}D^{\mathrm{Prop}}_{n, ap, t}$ considering the propagation delay for satellite communication with $D^{\mathrm{Prop}}_{n, ap, t} = d_{n, ap, t} / c$, where $D_{l}^{\mathrm{B}}$ is the specified delay bound for services in slice $l$ and $c$ is the light speed.

Due to the deterministic service rate, the effective capacity is defined as $C(R^{n, ap}_{l, t}, \theta) = -\frac{1}{\theta}\log\left(\mathbb{E}\left\{e^{-\theta R^{n, ap}_{l, t}}\right\}\right) = R^{n, ap}_{l, t}$.
Based on the effective capacity, with a given reserved resources from access point $ap$, the maximal supported intensity for the Poisson process can be given by \cite{chang2000performance}
\vspace{-1.5mm}
\begin{equation}
    \begin{aligned}
        \tilde{\lambda}_{n, ap, l, t} \!= \!\left\{ \!
        \begin{aligned}
            &\min\!\left\{\!\frac{\theta C(R^{n, ap}_{l, t}, \theta)}{e^\theta - 1}A_n, \lambda^\mathrm{C}_{n,l,t} \!\right\}\!, \!\!\!&\text{if } ap = 0, \\
            &\frac{\theta C(R^{n, ap}_{l, t}, \theta)}{e^\theta - 1}, &\text{otherwise},
        \end{aligned}
        \right.
    \end{aligned}
    \vspace{-1.5mm}
\end{equation}
where the maximum supported service demand from terrestrial networks is limited by $\lambda^\mathrm{C}_{n,l,t}$. 
To satisfy the service demand for each slice, the total supported service demand $\tilde{\lambda}^\mathrm{tot}_{n,l,t}$ should exceed the service demand for each slice, given by
\vspace{-1.5mm}
\begin{equation}
    \tilde{\lambda}^\mathrm{tot}_{n,l,t} = \sum_{ap \in \mathcal{AP}^\mathrm{ts}_{n,t}} \tilde{\lambda}_{n,ap,l,t} \geq \lambda_{n,l,t}. \label{op_demand}
    \vspace{-1mm}
\end{equation}

\vspace{-3.5mm}
\subsection{Problem Formulation}
\vspace{-0.5mm}

Considering the dynamic network environments with time-varying service demand and available resources for each cell, we aim to minimize the long-term overall system cost by reserving communication resources from different access points.
The overall system cost includes the resource cost from both terrestrial and satellite networks and the dissatisfaction cost for the delay-tolerant service. 
Specifically, the resource cost in time slot $t$ for cell $n$ is the sum of resource cost in different slices from different access points serving the target area, which can be given by $C^\mathrm{res}_{n, t} = \sum_{l}\sum_{ap \in \mathcal{AP}^\mathrm{ts}_{n,t}} \alpha_{ap} b^{n, ap}_{l, t} B_{ap}$,
where $\alpha_{ap}$ is the cost of a resource unit for access point $ap$.

Moreover, although there are no stringent delay requirements for the delay-tolerant service, each user may have an expected delay performance given by $D^\mathrm{B}_2$.
Let $W^\mathrm{dis}_{n, ap, l, t} = \mathrm{Pr}(D^{n,ap}_{l,t} > D_{n, ap, l, t}^{\mathrm{B}}) \mathbbm{1}_{b^{n, ap}_{l, t} > 0}$ denote the delay dissatisfaction probability indicating the probability of the delay exceeding the expectation for slice $l$ of cell $n$ in time slot $t$.
In this case, the dissatisfaction cost in time slot $t$ for cell $n$ can be represented by the maximum delay dissatisfaction probability from all access points serving users in the cell, given by
\vspace{-2mm}
\begin{equation}
    C^\mathrm{dis}_{n, t} =\max\nolimits_{ap \in \mathcal{AP}^\mathrm{ts}_{n,t}} W^\mathrm{dis}_{n, ap, 2, t} \ . \label{eq_dis}
    \vspace{-1.5mm}
\end{equation}

Let $\mathbf{b}_{n,ap,l} = \{b^{n, ap}_{l, 1}, ..., b^{n, ap}_{l, T}\}$ denote the vector of the resource reservation decision for slice $l$ in cell $n$ from access point $ap$ within $T$ time slots. 
To this end, we formulate the resource slicing problem in STIN to minimize the long-term overall system cost while satisfying delay and reliability requirements in the presence of network dynamics, given by
\vspace{-2.5mm}
{
\setlength{\belowdisplayskip}{0pt}
\setlength{\belowdisplayshortskip}{0pt}
\begin{align}
    \mbox{\textbf{P0} } \min_{\mathbf{b}_{n,ap,l}} \quad & \mathbb{E} \left[ \lim_{T \rightarrow \infty} \frac{1}{T}\sum_{t=1}^{T} \sum_{n \in \mathcal{N}} \left(\beta_1 C^\mathrm{res}_{n,t} + \beta_2 C^\mathrm{dis}_{n,t}\right)\right]  \label{optimal}\\[-0.25ex]
    \mbox{s.t.} \quad & W^\mathrm{dis}_{n, ap, 1, t} \leq \varepsilon, \tag{\ref{optimal}{a}} \label{op_a}\\[-0.25ex]
    & 0 \leq \sum_l b^{n, ap}_{l, t} \leq a_{n, ap, t}, \tag{\ref{optimal}{b}} \label{op_b}\\[-0.5ex]
    & \sum_{n} \sum_{l} b^{n, ap}_{l, t} \leq K, \forall ap \in \mathcal{S}, \tag{\ref{optimal}{c}} \label{op_c} \\[-0.5ex] 
    & \eqref{op_equal}, \eqref{op_demand}, \notag
    \vspace{-3mm}
\end{align} 
}where $\beta_1$ and $\beta_2$ represent the weights for two costs. 
Constraint \eqref{op_a} represents that the delay dissatisfaction probability should be less than $\varepsilon$ as the delay and reliability requirement of delay-sensitive services; constraint \eqref{op_b} indicates that for each access point, the overall ratio of reserved resources for different slices in one cell should be between 0 and 1 if the connectivity is accessible, otherwise, the ratio should be 0; and constraint \eqref{op_c} limits the maximum available beam resources from each satellite. \looseness=-1

\vspace{-1.5mm}
\section{Distributed Resource Slicing Scheme with Cross-Cell Coordination in STIN}
\vspace{-1mm}
\subsection{Problem Decomposition}
\vspace{-1mm}

Note that the formulated problem $\textbf{P0}$ is a non-linear stochastic optimization problem with spatiotemporal dynamic service demands and time-varying available resources in STIN. 
Therefore, conventional optimization methods face significant challenges in solving such long-term optimization problems with the space and time interactions among different cells. 
Furthermore, the continuous decision space for decisions from each access point for all slices makes the direct application of model-free techniques impractical.
Thus, we propose a distributed resource slicing (DRS) scheme to solve the problem.
Specifically, to reduce the decision dimension, considering that the resource reservation decisions remain unchanged within each slicing window, we decouple the original problem into the following two sub-problems: 
1) resource reservation problem with a given satellite set in each slicing window obtained from the second subproblem, and 2) satellite selection problem for the slicing window in each cell.

In the first subproblem, with a set of access points to serve the target cell, each controller determines the reservation ratio in each slicing window to minimize the system cost, given by 
\vspace{-5mm}
{
\setlength{\belowdisplayskip}{0pt}
\setlength{\belowdisplayshortskip}{0pt}
\begin{align}
    \mbox{\textbf{P1} } \min_{b^{n, ap}_{l, t}} \quad &  \sum_{t=t_{n,i}}^{t_{n,i} + w_{n,i}} \left(\beta_1 C^\mathrm{res}_{n,t} + \beta_2 C^\mathrm{dis}_{n,t}\right) \vspace{-1mm} \label{optimal1}\\[-0.5ex]
    \mbox{s.t.} \quad & ap \in \mathcal{AP}_{n,i}^{*}, \vspace{-1mm} \tag{\ref{optimal1}{a}} \label{op1_a} \\[-0.5ex]
    & \eqref{op_equal}, \eqref{op_demand}, \eqref{op_a}, \eqref{op_b}, \eqref{op_c}, \notag
    \vspace{-5mm}
\end{align}
}where $\mathcal{AP}_{n,i}^{*}$ is the selected access point set for $i$-th slicing window in cell $n$. 

In the second subproblem, we find the optimal set of access points to serve each cell in each slicing window for a long-term optimal overall system cost, which is given by
\vspace{-2mm}
\begin{align}
    \mbox{\textbf{P2} } \min_{\mathcal{AP}_{n,i}^{*}} \quad & \mathbb{E} \left[ \lim_{T \rightarrow \infty} \frac{1}{T}\sum_{t=1}^{T} \sum_{n \in \mathcal{N}} \left(\beta_1 C^\mathrm{res}_{n,t} + \beta_2 C^\mathrm{dis}_{n,t}\right)\right]  \label{optimal2}\\[-0.5ex]
    \mbox{s.t.} \quad & \mathcal{AP}_{n,i}^{*} \subseteq \mathcal{AP}_{n,i}, \tag{\ref{optimal2}{a}} \label{op2_a} \\[-0.5ex]
    & \eqref{op_equal}, \eqref{op_demand}, \eqref{op_a}, \eqref{op_b}, \eqref{op_c}. \notag
    \vspace{-1mm}
\end{align}

\vspace{-4mm}
\subsection{Distributed Optimization-Based Resource Reservation}
\vspace{-0.5mm}
In our considered STIN, one satellite can cover multiple terrestrial cells, leading to potential resource competition among different cells. 
Since different controllers independently make decisions without direct communication, it is hard to design a centralized optimization approach to find the optimal decisions for all cells to solve \textbf{P1}.
Therefore, we design a distributed optimization-based resource reservation algorithm
with centralized feedback from satellites, where the resource utilization of a satellite can be reported to its served cells to help the controller decisions.
The designed algorithm is implemented in a two-step process involving local solution derivation and global decision execution.

In the first step, each controller finds the local solution $[\tilde{b}^{n,ap}_{l,t}]^{|\mathcal{AP}_{n,i}^{*}| \times w_{n,i}}$ at the beginning of each slicing window to solve \textbf{P1} by neglecting constraint \eqref{op_c} for the potential resource competition.
Since the objective function \eqref{optimal1} and constraint \eqref{op_a} includes a conditional expression related to the variable $b^{n,ap}_{l,t}$, the problem cannot be directly solved by the optimization technique.
In this case, we add an additional binary variable $r^{n,ap}_{l,t}$ to constrain the resource reservation decision, equivalently transforming \textbf{P1} into
\vspace{-2mm}
{
\setlength{\belowdisplayskip}{-2pt}
\setlength{\belowdisplayshortskip}{-2pt}
\begin{align}
    \mbox{\textbf{P1'} } \min_{b^{n,ap}_{l,t}, r^{n,ap}_{l,t}} \quad &  \sum_{t=t_{n,i}}^{t_{n,i} + w_{n,i}} \beta_1 C^\mathrm{res'}_{n,t} + \beta_2 C^\mathrm{dis'}_{n,t}  \label{optimal1'}\\[-0.8ex]
    \mbox{s.t.} \quad & \mathrm{Pr}(D^{n,ap}_{1,t}\! > \! D_{n, ap, 1, t}^{\mathrm{B}}) \! + \! r^{n,ap}_{l,t}\! -\! 1\! <\! \varepsilon, \tag{\ref{optimal1'}{a}} \label{op1'_a} \\[-0.5ex]
    & b^{n,ap}_{l,t} \leq r^{n,ap}_{l,t}, \tag{\ref{optimal1'}{b}} \label{op1'_b}\\[-0.5ex]
    & r^{n,ap}_{l,t} \in \{0, 1\}, \tag{\ref{optimal1'}{c}} \label{op1'_c}\\[-0.5ex]
    & \eqref{op_equal}, \eqref{op_demand}, \eqref{op_a}, \eqref{op_b}, \eqref{op1_a}, \notag
\end{align}
}where
\vspace{-1mm}
{
\setlength{\belowdisplayskip}{0pt}
\setlength{\belowdisplayshortskip}{0pt}
\begin{align}
C^\mathrm{res'}_{n, t} &= \sum_{l}\sum_{ap \in \mathcal{AP}_{n,t}} \alpha_{ap} b^{n,ap}_{l,t} B_{ap} \omega_{ap}, \label{eq_res_new}\\[-0.8ex]
C^\mathrm{dis'}_{n,t}\!&=\!\! \max_{ap \in \mathcal{AP}_{n,t}}\!\!\mathrm{Pr}(D^{n,ap}_{2,t} > D_{n, ap, 2, t}^{\mathrm{B}}) + r^{n,ap}_{l,t} - 1. \label{eq_dis_new}
\end{align}
}For the first step, $\omega_{ap}$ remains 1, which is the parameter for the second step. 
Since the formulated mixed integer programming problem with conic constraints is a convex problem, the local solution for each cell $[\tilde{b}^{n,ap}_{l,t}]^{|\mathcal{AP}_{n,i}^{*}| \times w_{n,i}}$ can therefore be directly obtained via convex optimization solvers. 

Given local solutions, each controller then executes the decision for each access point.
Considering the potential exceeded utilization of the beam resources from each satellite, an iteratively distributed optimization algorithm is designed to reach the final feasible decision execution.
The main algorithm is given in Algorithm \ref{alg:IDOA} for the decision execution. 
With the local solution of each controller, when the access point becomes accessible within the time window, the controller will send local solutions to the corresponding access point for the execution.
If the local solutions overuse the satellite resources, the resource values will be reduced proportionally from the corresponding satellites as shown in lines \ref{alg:IDOA_s1}-\ref{alg:IDOA_e1}.
Otherwise, the reserved resource will remain the same within the slicing window during the satellite being accessible.
If the cumulative resource demands from all cells for a satellite exceeds the satellite's resource capacity, a penalty mechanism is introduced in the objective function to account for resource over-utilization.
Each controller needs to solve \textbf{P1'} until reaching the maximum number of iterations or the objective function value converges, given by line \ref{alg:IDOA_s2} to \ref{alg:IDOA_e2}.
Let $\omega_{ap}^{(u+1)} = (u \omega_{ap}^{(u)} + (u+1) \beta_3 \psi_{ap}^{(u)} / K ) / (u+1)$ represent the penalty of the resource for the satellite with exceeded utilization in ($u$+1)-th iteration, where $\psi_{ap}^{(u)}$ is the resource exceeding ratio and $\beta_3$ is the weight parameter.
Finally, the controllers execute decisions and update the local solutions.
\vspace{-5mm}
\begin{algorithm}
    \caption{Iteratively Distributed Optimization Algorithm}\label{alg:IDOA}
    \begin{algorithmic}[1]
    \setlength{\baselineskip}{0.9\baselineskip}
    \vspace{-1mm}
        \STATE \textbf{Initialization:} Local solution $\tilde{b}^{n,ap}_{l,t}$;
        \WHILE{$u < iter_{\max}$ and not converge}
            \FOR {$ap \in \mathcal{S}_{n,t}$} \label{alg:IDOA_s1}
                \IF{$\tilde{b}^{n,ap}_{l,t}$ not executed and $a^{n,ap}_{l,t} = 1$}
                    \STATE Send the local solution to the satellites;
                    \STATE Receive utilization feedback and proportionally reduced resource values if exceeding the capacity; \label{alg:IDOA_e1}
                \ENDIF
            \ENDFOR
            \IF{$\exists$ Eq. \eqref{op_c} not satisfied}
                \STATE Solving \textbf{P1'} and update $\tilde{b}^{n,ap}_{l,t}$, $\omega_{ap}^{(u)}$; \label{alg:IDOA_s2}
                \STATE Go back to line \ref{alg:IDOA_s1}; \label{alg:IDOA_e2}
            \ENDIF
        \ENDWHILE
        \STATE Execute decisions to $b^{n,ap}_{l,t}$ and update $\tilde{b}^{n,ap}_{l,t}$;
        \STATE \textbf{Output:} $\tilde{b}^{n,ap}_{l,t}$ and $b^{n,ap}_{l,t}$;
    \end{algorithmic}
    \vspace{-1mm}
\end{algorithm}

\vspace{-5mm}
\subsection{Reinforcement Learning-Based Satellite Selection}
\vspace{-0.5mm}

Next, we will investigate the optimal satellite selection for each cell in each slicing window. 
Since the formulated satellite selection problem is an integer nonlinear optimization problem with asynchronous decision-making from each controller, it is complex to obtain optimal decisions by deterministic optimization algorithms. 
Thus, we reformulate problem \textbf{P2} as a decentralized partially observable Semi-Markov decision process (Dec-POSMDP) \cite{feriani2021single} to model the time-varying service demand and interaction among different cells.

We model the Dec-POSMDP as a tuple $(\mathcal{N}, \mathfrak{S}, \underline{\mathfrak{A}}, \mathcal{P}, \underline{\mathfrak{R}}^w, \mathcal{W}, \gamma, \underline{\mathcal{O}}, Z)$, including the agent space $\mathcal{N}$ represented by the set of controllers in each cell, the state space $\mathfrak{S}$, the action space of each controller $\underline{\mathfrak{A}}$, the state transition function $\mathcal{P}:= \mathfrak{S} \times \underline{\mathfrak{A}} \times \mathfrak{S} \to [0,1]$, the reward function for each slicing window $\mathfrak{R}^w:= \mathfrak{S} \times \underline{\mathfrak{A}} \times \mathcal{W} \to \mathbb{R}$, slicing window space $\mathcal{W}$, discounted factor $\gamma$, observation spaces from each controller $\underline{\mathcal{O}}$, and observation function $\mathcal{Z}:= \mathfrak{S} \times \underline{\mathfrak{A}} \times \underline{\mathcal{O}} \to [0,1]$, where $\underline{\,\cdot\,}$ represents the joint parameters for all controllers.
Let $\zeta_{n}^i = [\lambda_{n, l, t'}, POS_{ap, t'}, a_{n, ap, t'}, w_{n, i}, \tilde{b}^{n,ap}_{l,t_{n,i}}]_{ap \in \mathcal{AP}_{n,i}, t' \in \mathcal{W}_{n}^i} \in \underline{\mathcal{O}}$ denote the observation of the controller in cell $n$ in $i$-th slicing window, where $POS_{ap, t'}$ is the position of access point $ap$. 
The state in time slot $t$ is the joint observation of each controller, given by $\Xi_t = \bigcup_{\{\{n, i\} | t \in \mathcal{W}_{n}^i\}} \zeta_{n}^i$. 
The action in slicing window $i$ for the controller in cell $n$ is denoted by $\mathcal{A}_{n}^{i} =  \mathcal{AP}_{n, i}^{*} \in \mathfrak{A}$ for the access point selection, which is determined at the beginning of each slicing window in each cell.
To evaluate an action in a state, we define the obtained reward for each controller in each slicing window as
\vspace{-2mm}
\begin{equation}
    \mathcal{R}_{n,i}^w(\zeta_{n}^i, \mathcal{A}_{n}^i, \mathcal{W}_{n}^i) = \sum\nolimits_{t \in \mathcal{W}_{n}^i} \gamma^{t - t_{n, i}} (-C^\mathrm{sys}_{n, t} - V^\mathrm{pen}_{n,t}),
    \vspace{-2mm}
\end{equation}
where $V_{n,t}$ is the penalty value indicating the violation of constraints \eqref{op_demand} and \eqref{op_a}, given by 
\vspace{-2mm}
\begin{equation}
    \hspace{-2mm}\resizebox{0.92\linewidth}{!}{$V^\mathrm{pen}_{n,t} = p_1 \sum\limits_{l}\mathbbm{1}_{\tilde{\lambda}^\mathrm{tot}_{n,l,t} < \lambda_{n,l,t}} + p_2 \max\limits_{ap \in \mathcal{AP}}\{\mathbbm{1}_{W^\mathrm{dis}_{n, ap, 1, t} > \varepsilon}\}$}
    \vspace{-2.5mm}
\end{equation}
with the penalty weights $p_1$ and $p_2$, and $\gamma \in (0, 1)$ reflects the importance of future reward compared with immediate reward.
Let $\mathbf{\Pi}$ denote the set of all policies.
To estimate the expected return of the current decision-making, a discounted cumulative reward function is considered and the objective function is transformed as
\vspace{-1.5mm}
\begin{equation}
    \max_{\pi \in \Pi} \mathbb{E} \left[\sum_{i'= 0}^\infty \sum_{n \in \mathcal{N}}\gamma^{t_{n, i'}} \mathcal{R}_{n,i}^w(\zeta_{n}^i, \mathcal{A}_{n}^i, \mathcal{W}_{n}^i) \bigg| \mathbf{\pi}\right].
    \vspace{-1.5mm}
\end{equation}

To solve the problem, we propose an asynchronous multi-agent proximal policy optimization (AMAPPO)-based algorithm \cite{yu2023asynchronous}, where each agent can take actions asynchronously. A centralized training and decentralized execution (CTDE) paradigm is adopted, where each controller makes decisions based on individual observations and updates the policy from the global state in a centralized manner. For each controller, the actor network parameterized by $\phi$ is responsible for the selection of access points for each slicing window in its cell given the local observation. The critic network parameterized by $\vartheta$ is leveraged to approximate a state-value function by generating the value of the given global state for updating the actor networks. Due to multiple satellites that can serve a cell in a slicing window, a large action dimension is introduced for selecting multiple satellites from a controller. Moreover, considering the mobility of LEO satellites, the accessible satellites for each controller in different slicing windows may be different. In this case, the action can be presented as a multi-discrete action set \cite{kanervisto2020action}, given by $\mathcal{A}_{n}^i = \{\mathcal{A}_{n, ap}^{i} \in \{0, 1\} \ |\  ap \in \mathcal{AP}_{n,i}\}$, with the corresponding mask to avoid some inaccessible access points being selected, where $\mathcal{A}_{n, ap}^{i} = 1$ represents the access point $ap$ will serve cell $n$ in $i$-th slicing window.

For the AMAPPO-based algorithm, the goal is to find the optimal parameter $\phi$ for the policy that maximizes the following clipped surrogate objective \cite{schulman2017proximal}
\vspace{-1mm}
\begin{equation}
\hspace{-1.5mm}
    L^\mathrm{clip}(\phi) \! = \! \mathbb{E} \! \left\{\min(\xi_{\phi} A^{\pi_{\phi_\mathrm{old}}} \!,\! \mathrm{clip}(\xi_{\phi},\! 1\!-\!\delta, \! 1\!+\!\delta)A^{\pi_{\phi_\mathrm{old}}})\right\},
    \vspace{-1mm}
\end{equation}
where $\xi_{\phi} = \pi_{\phi}(\mathcal{A}|\zeta) / \pi_{\phi_\mathrm{old}}(\mathcal{A}|\zeta)$ is the policy updating ratio based on current policy for $\phi$.
Moreover, $A^{\pi_{\phi_\mathrm{old}}}$ is the estimated advantage from policy $\pi_{\phi_\mathrm{old}}$, which can be expressed as $A^{\pi_{\phi_\mathrm{old}}}(\zeta_{n}^i, \mathcal{A}_{n}^i, \mathcal{W}_{n}^i) = \mathcal{R}_{n,i}^w(\zeta_{n}^i, \mathcal{A}_{n}^i, \mathcal{W}_{n}^i) \!+\! \gamma^{\Delta t^{i+1}_{n, i}} V_\vartheta(\Xi_{t_{n, i+1}}) \!-\! V_\vartheta(\Xi_{t_{n, i}})$, where $\Delta t^{i+1}_{n, i} = t_{n, i+1} - t_{n, i}$.
The state-value function of the critic network is given by
\vspace{-1mm}
\begin{equation}
\hspace{-2mm}
    \resizebox{0.92\linewidth}{!}{$V_\vartheta(\Xi_{t_{n, i}}) = \mathbb{E}\!\left\{ \! \sum_{k=0}^\infty \! \gamma^{\Delta t^{i+k}_{n, i}} \! \mathcal{R}_{n,i}^w(\zeta_{n}^{i+1+k}, \! \mathcal{A}_{n}^{i+1+k}, \! \mathcal{W}_{n}^{i+1+k}) \,|\, \Xi_{t_{n, i}}, \! n  \right\} \!$ .}
    \vspace{-1mm}
\end{equation}
The clipping function clips the ratio $\xi_\phi$ between $1-\delta$ and $1 + \delta$ with a clipping hyperparameter $\delta$, which removes incentives for the new policy to get far from the old policy to improve the learning stability. 

The overall AMAPPO-based resource slicing algorithm is given in Algorithm \ref{alg:AMAPPO}. 
The buffer $\mathcal{B}$ is used to store the transitions of the interaction with the environment. 
As shown from lines \ref{AMAPPO_s1}-\ref{AMAPPO_e1}, at the beginning of each slicing window, the controller of each cell obtains the observation from the environment and takes actions based on the current policy. 
At the end of each slicing window, the controller calculates the reward and records transition data into the buffer, including the state, observation, action, reward of a slicing window, and the next state.
At the end of each episode, the policy and value function will be updated based on the clipped surrogate objective and the mean-square error based on lines \ref{AMAPPO_s2}-\ref{AMAPPO_e2}.

\vspace{-2mm}
{
\setlength{\belowdisplayskip}{-2pt}
\setlength{\belowdisplayshortskip}{-2pt}
\begin{algorithm}
    \caption{AMAPPO-Based Satellite Selection algorithm}\label{alg:AMAPPO}
    \begin{algorithmic}[1]
    \setlength{\baselineskip}{0.88\baselineskip}
    \vspace{-1mm}
        \STATE \textbf{Initialization:} Parameters $\phi$, $\vartheta$, and buffer $\mathcal{B} = \{\}$;
        \WHILE{$episode \leq episode_{\max}$}
            \FOR{$t \in \{1,2,...,T\}$}
                \STATE Obtain the state $\Xi_{t}$;
                \FOR{$n \in \mathcal{N}$} \label{AMAPPO_s1}
                    \IF{$t = t_{n, i}$}
                        \STATE Obtain the observation $\zeta_{n}^i$ and action of access point selection $\mathcal{A}_{n}^i$ based on the policy $\pi_\phi$; 
                        \STATE Solving \textbf{P1'} to obtain the initial local solution;
                    \ENDIF
                \ENDFOR
                \STATE Apply Algorithm \ref{alg:IDOA} for reserving resources;
                \FOR{$n \in \mathcal{N}$}
                    \IF{$t = t_{n, i} + w_{n, i} - 1$}
                        \STATE Compute reward $\mathcal{R}_{n,i}^w(\zeta_{n}^i, \mathcal{A}_{n}^i, \mathcal{W}_{n}^i)$ and insert transition data into buffer $\mathcal{B}$;  \label{AMAPPO_e1}
                    \ENDIF
                \ENDFOR
            \ENDFOR
            \STATE Update the policy parameters $\phi \! \gets \! \arg\max_{\phi^\prime}\! L^\mathrm{clip}(\phi^\prime)$; \label{AMAPPO_s2}
            \STATE Update critic network parameter \\ $\vartheta \gets \arg\min_{\vartheta^\prime}(\mathbb{E} \! \left[\mathcal{R}_{n,i}^w \!+\! V_\vartheta(\Xi_{t_{n, i+1}}) \!- \! V_\vartheta(\Xi_{t_{n, i}})\right])^2$ \label{AMAPPO_e2}
            \STATE Empty buffer $\mathcal{B}$;
        \ENDWHILE
        \STATE \textbf{Output:} Parameter $\phi$;
    \end{algorithmic}
\end{algorithm}
}

\vspace{-4mm}
\section{Simulation Results}
\vspace{-1mm}

In this section, we present simulation results to demonstrate the performance of our proposed DRS scheme in STIN.
We adopt the Starlink Phase 1 for the satellite constellation parameters, including 72 orbits with 22 satellites on each orbit. The inclination angle is 53 degrees and the altitude of the satellite is 550 km. By default, we set the elevation angle to be 40 degrees.
Each satellite has 3 beams to serve different cells simultaneously.
The satellite transmit power is set to 10 dBW, with an antenna gain of 50 dBi and pathloss factor of 2.5.
The transmit power of base stations is 32 dBm with an antenna gain of 10 dBi and pathloss factor of 3.5. 
In the simulation, we consider 7 cells in total, where each cell has different numbers of base stations and diversified service demand.
The overall time of the simulation includes 300 time slots where the duration of each time slot is set to 20 s.
The maximum slicing window length is 300 s.
The delay and reliability requirement for delay-sensitive applications is set to 50 ms with $\varepsilon = 0.01$ and the packet size of 0.2 Mb, while the expected delay for delay-tolerant applications is 0.3 s with packet size of 2 Mb.

To evaluate the effectiveness of the proposed DRS scheme, we adopt the following two benchmark algorithms: 1) \textit{Pure-AMAPPO}: the resource reservation decisions are directly made by AMAPPO-based algorithm by outputting the ratio of resources reserved for each slice from each access point, adopting the same learning rate of the proposed DRS scheme; and 2) \textit{IDOA}: the resource reservation decisions are obtained from Algorithm \ref{alg:IDOA}, where resources from all satellites covering the cell within the slicing window can be reserved.

\begin{figure}[t]
    \centering
    \begin{minipage}{.25\textwidth}
        \centering
        \hspace{-5mm}
        \includegraphics[width=.87\linewidth]{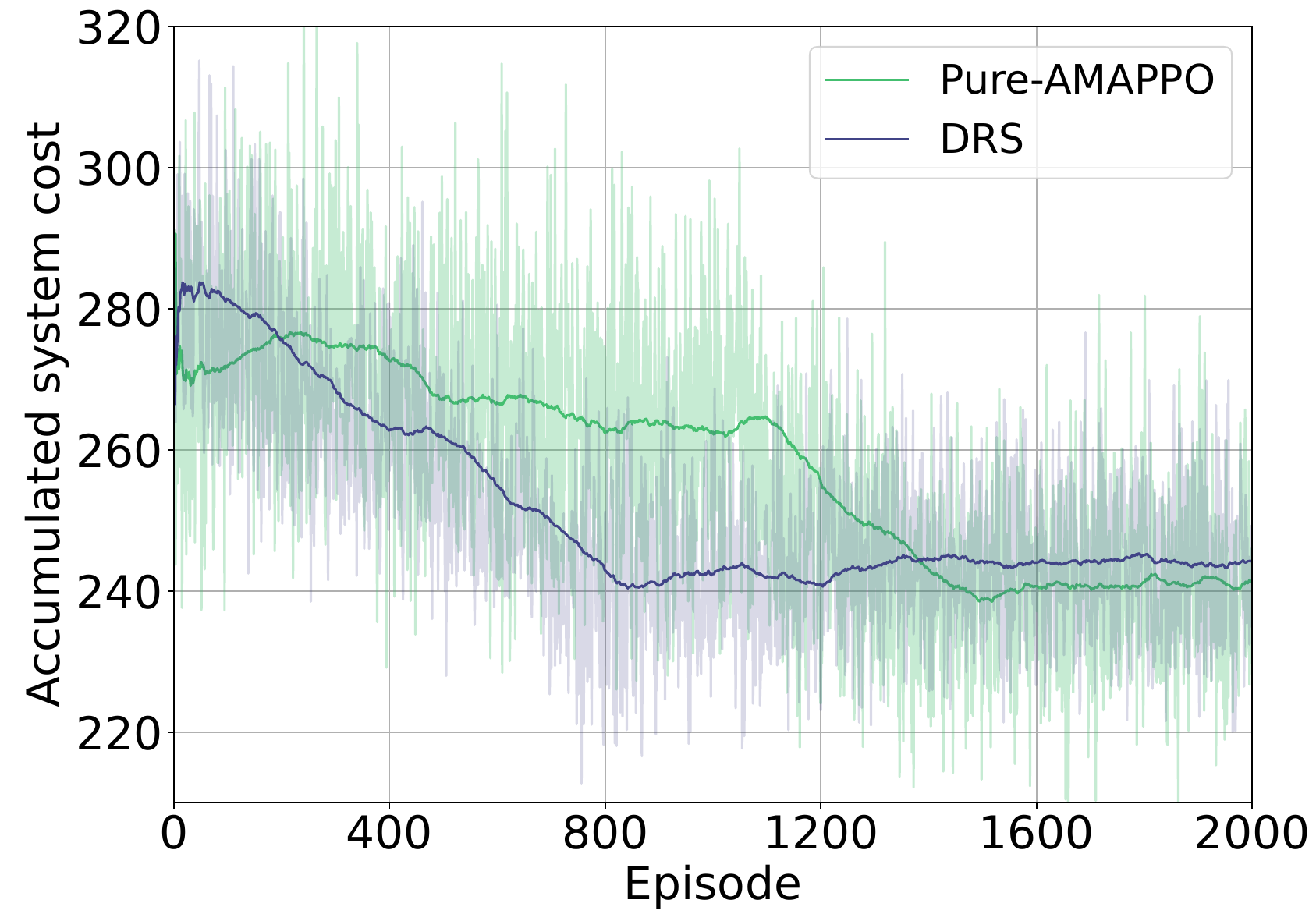}
        \vspace{-1.7mm}
        \caption{\hspace{-0.5mm} Convergence performance.}
        \label{fig:convergence}
    \end{minipage}%
    \begin{minipage}{.25\textwidth}
        \centering
        \hspace{-5mm}
        \includegraphics[width=0.99\linewidth]{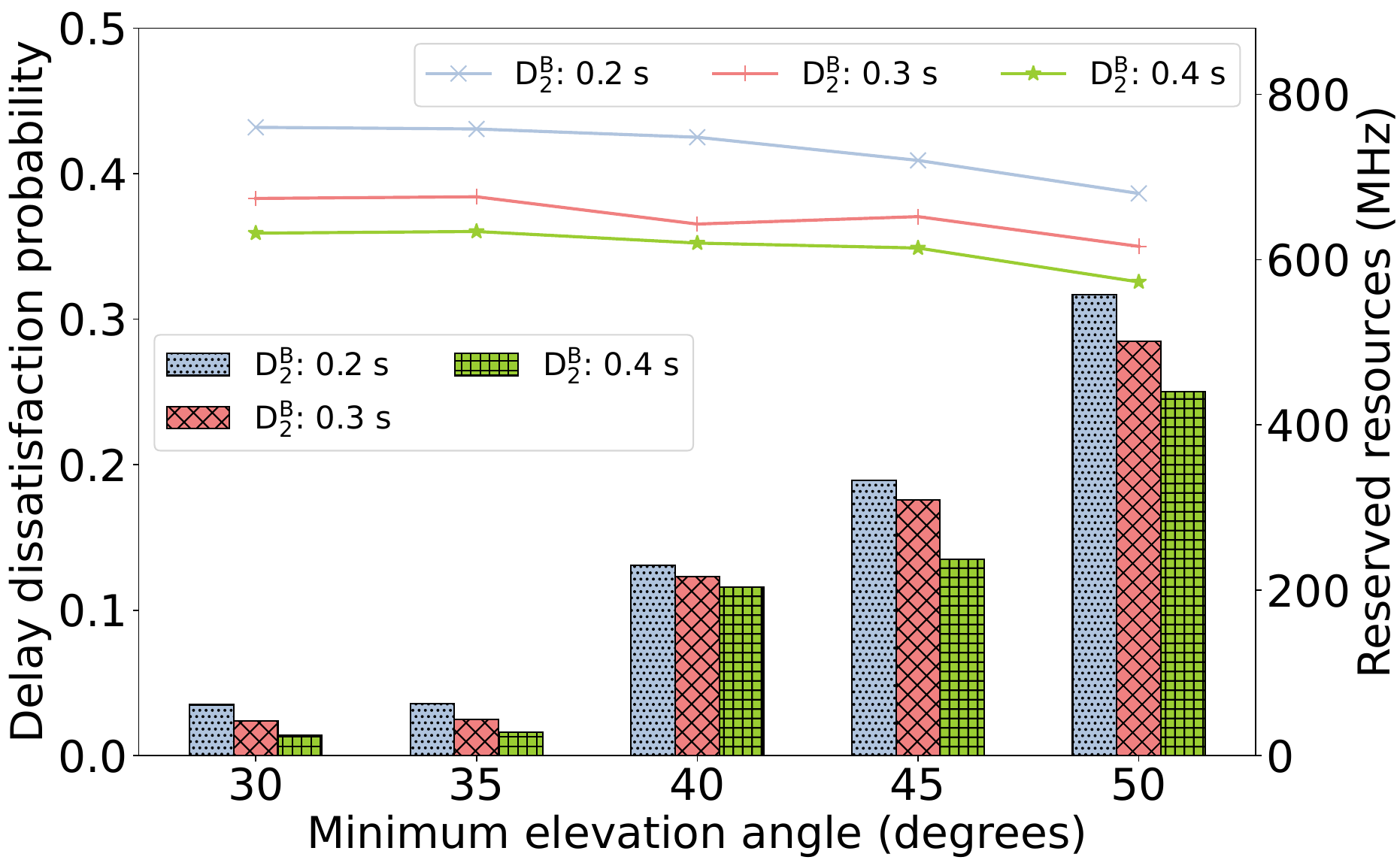}
        \vspace{-1.7mm}
        \caption{\hspace{-0.5mm}Elevation angles impacts.}
        \label{fig:elevation}
    \end{minipage}
    \vspace{-3mm}
\end{figure}

\begin{figure}[t]
\vspace{-1.5mm}
	\centering
	\subfigure[Same service demand]{\includegraphics[width=0.23\textwidth]{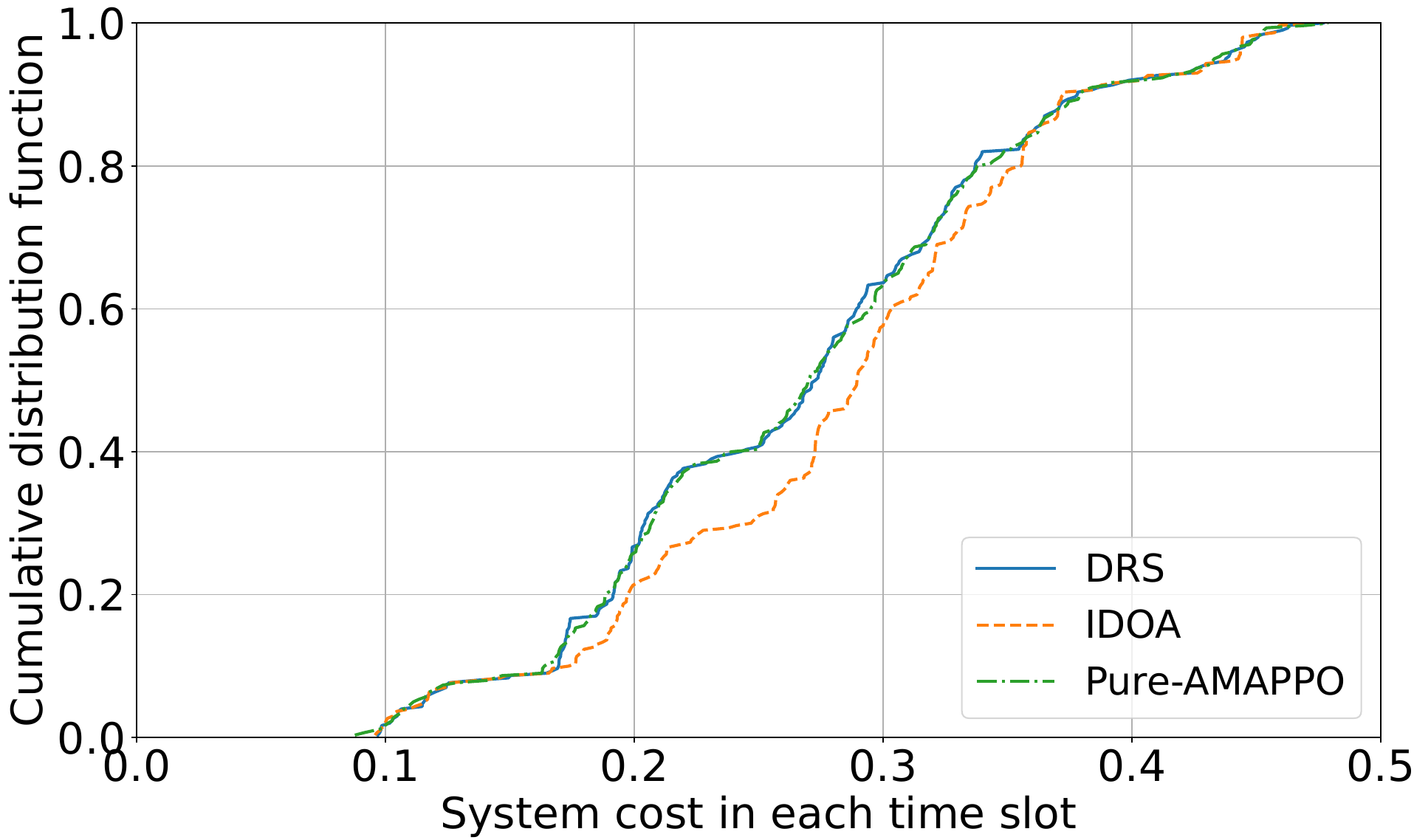}\label{fig:sync}}
	\subfigure[Different service demand]{\includegraphics[width=0.225\textwidth]{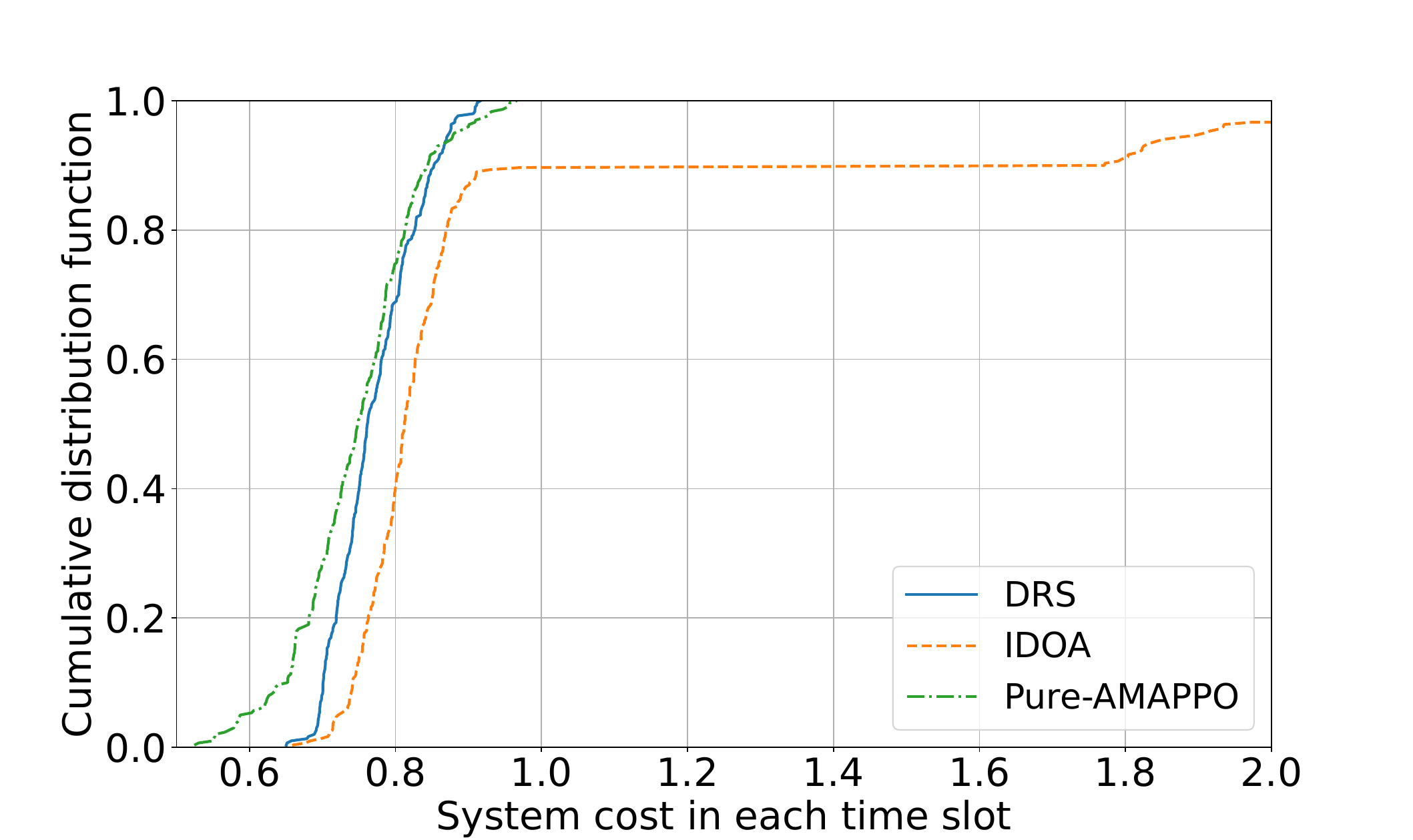}\label{fig:async}}
 \vspace{-3mm}
	\caption{\small CDF of system cost in each time slot.}
	\label{fig:demand}
 \vspace{-1mm}
\end{figure}

We first present the convergence performance of the overall system cost in each iteration of the proposed DRS scheme during the training period, compared with the \textit{Pure-AMAPPO} algorithm, as shown in Fig. \ref{fig:convergence}. We conducted the simulation with 2000 iterations. 
It can be seen that the accumulated system cost in each episode of our proposed DRS scheme converges after around 850 episodes, which has a faster convergence speed compared with the \textit{Pure-AMAPPO} which converges after around 1500 episodes.
This indicates that our proposed scheme in a hybrid data-model co-driven approach can achieve optimal results with a short training time.

Next, we show the resource consumption performance and delay dissatisfaction probability of our proposed scheme in different expected delays of delay-tolerant services under different minimum elevation angles in Fig. \ref{fig:elevation}. 
From the figure, we can see that when the elevation angle is small, indicating that more satellites are able to serve users in each cell, a large amount of resources are reserved to ensure a small delay dissatisfaction probability, which enables a small overall system cost with the optimal tradeoff for delay-tolerant applications. 
With the increment of the elevation angle, the resource decreases and the delay dissatisfaction probability increases correspondingly. 
Specifically, some unavoidable satellite resource competition from different cells or even a service disruption from satellite networks with a large elevation angle may result in a large delay dissatisfaction probability.

Finally, as shown in Fig. \ref{fig:demand}, we compare the time slot-averaged system cost with benchmark algorithms in two scenarios.
The first scenario assumes that all cells have identical service demands where the slicing window lengths are the same in every cell.
The second scenario considers varying service demands with asynchronous resource reservation decisions in each cell.
From Fig. \ref{fig:sync}, we can see that our proposed DRS scheme outperforms the \textit{IDOA} and has similar performance to the \textit{Pure-AMAPPO} since the reinforcement learning-based satellite selection can effectively reduce the competition among different cells. 
With the different service demands in each cell, as shown in Fig. \ref{fig:async}, the performance of the DRS scheme can approach the \textit{Pure-AMAPPO} and outperform the \textit{IDOA}, indicating that our proposed scheme can effectively take the temporal and spatial interactions among different cells into account to reach a lower system cost.

\vspace{-0.8mm}
\section{Conclusion}
\vspace{-0.8mm}

In this paper, we have investigated resource slicing with cross-cell coordination in STIN to minimize the system cost in terms of resource usage and delay performance for different services.
To cope with the spatiotemporal dynamics of service demand and satellite mobility, which result in asynchronous slicing decisions and resource competition across different cells, we have proposed a DRS scheme by developing a hybrid data-model co-driven approach for resource slicing in STIN, which can effectively reduce the training time compared with a fully data-driven method.
Specifically, an AMAPPO-based algorithm is designed to find the optimal satellite set for each cell, and a distributed optimization-based algorithm is introduced for resource reservation to find the minimum cost in each slicing window. 
In the future, we will explore the collaboration among different satellite beams for different cells with adjustable beam resources to adaptively manage resources for various slices in STIN.\looseness=-1

\vspace{-0.5mm}

\bibliographystyle{IEEEtran}
\bibliography{Bibliography}

\end{document}